\documentclass[aps, prab, reprint, superscriptaddress, twocolumn, amsmath, amssymb, amsfonts, floatfix, longbibliography]{revtex4-2}

\usepackage{mathrsfs}
\usepackage{graphicx}
\usepackage{dcolumn}
\usepackage{bm}
\usepackage[breaklinks=true, pdftitle={}]{hyperref}
\usepackage[normalem]{ulem}
\usepackage{epstopdf}
\usepackage{xcolor}
\usepackage[separate-uncertainty]{siunitx}

\newcommand{\bl}[1]{{\color{black}#1}}

\newcommand{\bd}[1]{\boldsymbol{#1}}

\begin{document}

\title{Spin-polarized electron beam generation in the colliding-pulse injection scheme}

\author{Zheng Gong}\email{gong@mpi-hd.mpg.de}
\affiliation{Max-Planck-Institut f\"ur Kernphysik, Saupfercheckweg 1, 69117 Heidelberg, Germany}
\author{Michael J. Quin}
\affiliation{Max-Planck-Institut f\"ur Kernphysik, Saupfercheckweg 1, 69117 Heidelberg, Germany}
\author{Simon Bohlen}
\affiliation{Deutsches Elektronen-Synchrotron DESY, Notkestr. 85, 22607 Hamburg, Germany}
\author{Christoph H. Keitel}
\affiliation{Max-Planck-Institut f\"ur Kernphysik, Saupfercheckweg 1, 69117 Heidelberg, Germany}
\author{Kristjan P\~oder}
\affiliation{Deutsches Elektronen-Synchrotron DESY, Notkestr. 85, 22607 Hamburg, Germany}
\author{Matteo Tamburini}\email{matteo.tamburini@mpi-hd.mpg.de}
\affiliation{Max-Planck-Institut f\"ur Kernphysik, Saupfercheckweg 1, 69117 Heidelberg, Germany}

\date{\today}

\begin{abstract}
Employing colliding-pulse injection has been shown to enable high-quality electron beams to be generated from laser-plasma accelerators. Here by leveraging test particle simulations, Hamiltonian analysis, and multidimensional particle-in-cell (PIC) simulations, we lay the theoretical framework of spin-polarized electron beam generation in the colliding-pulse injection scheme. Furthermore, we show that this scheme enables the production of quasi-monoenergetic electron beams in excess of 80\% polarization and tens pC charge with commercial 10-TW-class laser systems.
\end{abstract}

\maketitle

\section{\label{section I} Introduction}

Particle accelerators are broadly used in material science~\cite{lee2018accelerator}, biology~\cite{karzmark1960technique}, medicine~\cite{qaim2021continuing}, fusion research~\cite{hofmann2018review}, industry~\cite{moller2020accelerator}, and as sources of intense and energetic photons~\cite{Godwin1969, deacon1977first, corde2013femtosecond, yan2017high, hussein2019laser, kettle2019single, Wang2021}. In science, one of the most important roles of accelerators is to probe the properties of fundamental forces as well as particles' structure in searches of possible physics beyond the standard model~\cite{bezrukov2012higgs}. Conventional accelerators have an accelerating gradient limit around $100\,\mathrm{MV/m}$ owing to the electrical breakdown of radio-frequency cavities. By contrast, laser-plasma-based accelerators can support accelerating fields above $100\,\mathrm{GV/m}$~\cite{tajima1979, esarey_LWFA_RMP}, enabling acceleration of electron beams to several GeV energy on centimeter scales~\cite{gonsalves2019petawatt}. Several experiments demonstrated the efficacy and robustness of the laser-wakefield acceleration (LWFA) mechanism~\cite{faure2004laser, geddes2004high, mangles2004monoenergetic, wang2013quasi, leemans2014multi, gonsalves2019petawatt, MaierPRX20, BohlenPRAB22}. Compared with conventional large-scale accelerators, plasma-based accelerators generally have advantages in costs, size, and achievable peak current. Thus, LWFA is regarded as a promising route to realizing compact lepton colliders~\cite{Leemans2009, Steinke2016, gschwendtner2019plasma}. 

To enable LWFA-based spin-dependent process investigations, which could also benefit high-energy lepton colliders~\cite{moortgat2008polarized, shiltsev2021modern}, it is crucial to develop all-optical methods for the controlled and reliable generation of highly polarized electron beams. Recently, theoretical schemes based on the collision of an ultrarelativistic electron beam with a tailored laser pulse have been proposed as a possible source of spin-polarized electron beams induced by hard-photon emissions in the strong-field QED regime~\cite{Seipt_2018, li2019ultrarelativistic, Chen_2019, Seipt_2019, geng2020spin, Xue2022} or by helicity transfer~\cite{li2022helicity}. However, the above methods require high-power and high-intensity laser pulses and are unsuitable for operating at a high-repetition rate.

In order to generate a high-current spin-polarized electron beam, Wen \textit{et al.} have put forward a scheme based on the laser-wakefield acceleration of pre-polarized plasma electrons with a density down ramp for injection~\cite{wen2019polarized}. With 3D particle-in-cell (PIC) simulations, this method was shown to deliver 0.31~kA electron beams with 90.6\% spin polarization by using a $2.1 \times 10^{18}\text{ W/cm}^2$ tens femtoseconds laser pulse with approximately 2.2~TW power~\cite{wen2019polarized}. In the scheme by Wen \textit{et al.}, \bl{a pre-polarized plasma is first produced via laser-induced molecular photodissociation, a method successfully employed in experiments to generate a high-density electron-spin-polarized gas with densities from $10^{16}\mathrm{cm}^{-3}$ to $10^{19}\mathrm{cm}^{-3}$~\cite{rakitzis2003spin_science, sofikitis2017spin_PRL, sofikitis2018spin_PRL}. Note that, in practice, it is not the entire plasma source that must be pre-polarized, but rather only the restricted injection volume itself. 
Although the pre-polarized plasma lifetime is of the order of 10 ns, hyperfine coupling results in a periodic electron-to-nucleus spin transfer with $\sim100$ ps period~\cite{sofikitis2017spin_PRL}. Thus, the driving laser pulse can arrive tens of picoseconds after plasma pre-polarization, which is easily achievable with existing laser technology.
Furthermore, while the alignment of laser pulses inside a plasma source with micrometer precision in both space and time has been demonstrated (see, e.g., Ref.~\cite{bohlenPRL22}), the laser pulses employed for plasma pre-polarization have low intensity requirements, such that their focal size can be large enough to easily enable spatial overlap. Remarkably, with the advent of 100~nm lasers~\cite{drescherNP21}, the molecular photodissociation technique might be applied to pure hydrogen, potentially enabling 100\% plasma pre-polarization.} \bl{The method of plasma pre-polarization via laser-induced molecular photodissociation was initially proposed by Hutzen \textit{et al.} and applied to polarized proton beam generation in laser-plasma interaction in Ref.~\cite{H2019Polarized}. Following these seminal works, other schemes leveraging a pre-polarized plasma were \bl{put forward} to generate energetic spin-polarized electron (or proton) beams~\cite{wu2019polarized_NJP, wu2019polarized_PRE, Buscher_2019, jin2020spin, gong2020energetic, thomas2020scaling, li2021polarized, reichwein2021robustness, reichwein2022acceleration, reichwein2022spin, reichwein2022particle, Fan_2022_njp, yan2022generation} or to investigate polarization effects in inertial confinement fusion~\cite{Hu_2020_PRE}.} More recently, Nie \textit{et al.} proposed to exploit the spin-dependent ionization cross section of xenon atoms to generate up to $\sim 31\%$ spin-polarized and 0.8~kA current electron beams in a beam-driven plasma wakefield accelerator~\cite{nie2021situ}. However, the above methods have limitations in the attainable charge or spin polarization of the beam, and no simple route exists to control the generated beam features. 

\begin{figure}[tb]
\includegraphics[width=1\columnwidth]{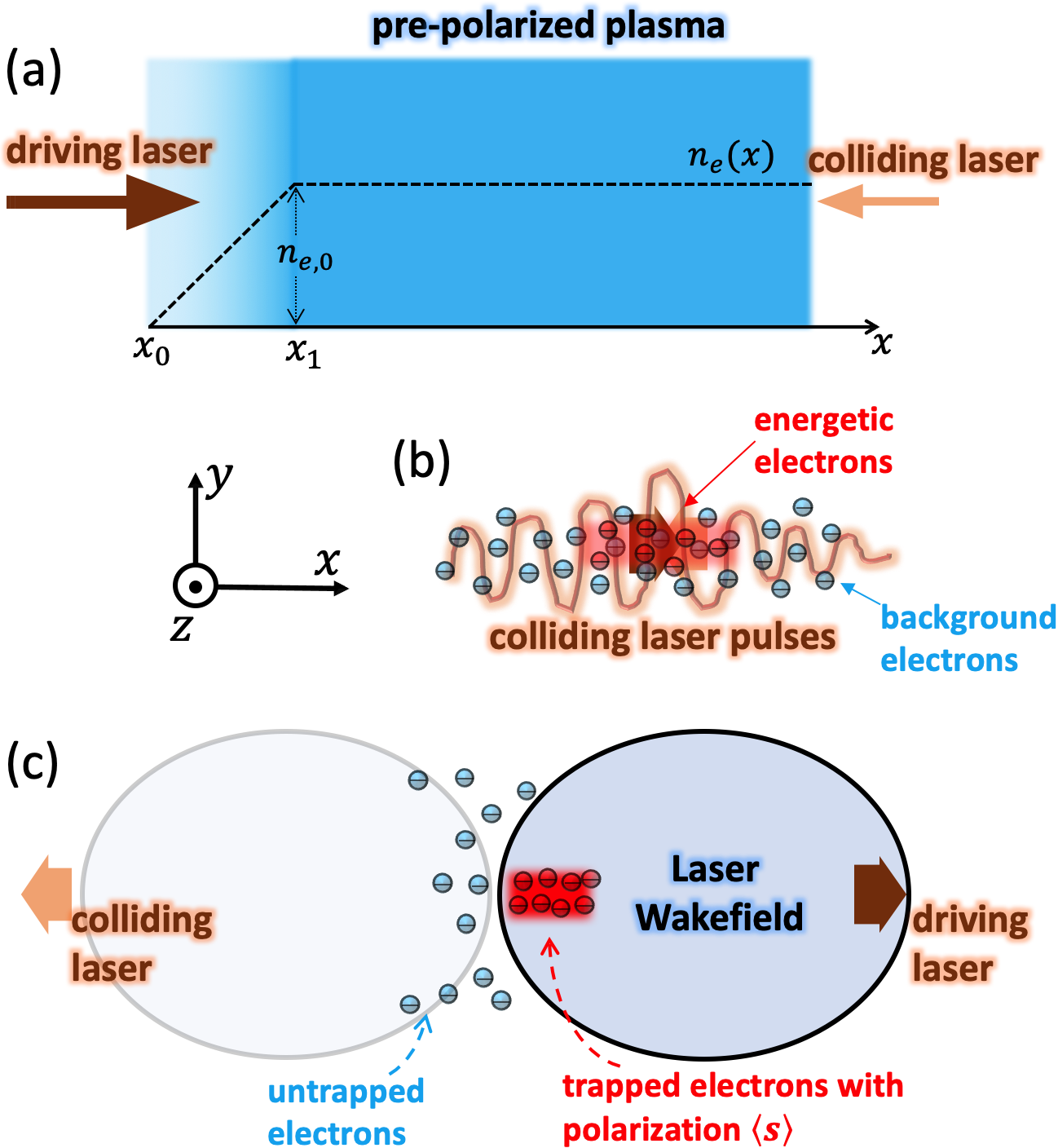}
\caption{Schematic of colliding pulse injection. \bl{(a) Two colliding laser pulses irradiate a pre-polarized underdense plasma with longitudinal density profile $n_e(x)$ shown by the black dashed line.} (b) Some plasma electrons (blue) undergo collisionless heating and gain residual energy and longitudinal momentum (red). (c) The electrons that have gained sufficient longitudinal momentum (red) to satisfy the injection criterion are trapped and subsequently accelerated in the wakefield.} \label{fig:schematic}
\end{figure}

The colliding-pulse injection (CPI) scheme~\cite{umstadterPRL96, esarey1997electron} has produced high-quality electron beams of low divergence and energy spread~\cite{faureN06, kotaki2009electron, rechatin2009controlling, lundh2011few}, which are stable and reproducible~\cite{faureN06, davoine2009cold, malkaPoP09, kotaki2009electron, hanssonNIMA16}. CPI provides many degrees of freedom that permit  control over the generated beam features. For instance, the produced electron beam energy, charge and energy spread are tunable by adjusting the position of the collision point in the plasma source~\cite{faureN06}, or the relative polarization between the driving and colliding laser pulses~\cite{rechatin2009controlling}. This renders CPI a robust and versatile alternative to single pulse LWFA to reliably generate high-quality, high-current spin-polarized electron bunches \bl{from a pre-polarized plasma}, as shown below (see also Ref.~\onlinecite{bohlenInPrep}). In CPI, a driving laser pulse with relativistic intensity induces a wakefield, while a sub-relativistic-intensity colliding pulse enables injection in the wake. As schematically illustrated in Fig.~\ref{fig:schematic}, the interaction process consists of two stages: (i) stochastic collisionless heating of plasma electrons by the two colliding laser pulses; (ii) trapping and acceleration of some energized electrons by the wakefield excited by the driving laser pulse.

In this article, we develop a model with an effective Hamiltonian that characterizes the electron dynamics, validate its predictions against quasi-3D PIC simulations, and show that CPI enables the generation of high-current and highly spin-polarized electron beams with controllable average spin polarization. The optimization of the beam features, including its charge and spin polarization, is studied in more detail in Ref.~\cite{bohlenInPrep}. Our work is arranged into four sections. Section~\ref{section II} presents 2D PIC simulation results showing the plasma electron dynamics with and without the colliding laser pulse. In section~\ref{section III}, we discuss electron heating and the injection criterion. In section~\ref{section IV}, quasi-3D PIC simulations assuming near cylindrical symmetry are performed to validate the theoretical model and further elucidate the electron injection dynamics and its influence on the charge and polarization of the electron beam. Our results are summarized in section~\ref{section V}, while the details of the particle spin pusher that we implemented in the spectral numerical-dispersion-free quasi-3D PIC code FBPIC~\cite{lehe2016spectral} are detailed in the Appendix.

\section{\label{section II} 2D simulations}

\begin{figure}[tb]
\includegraphics[width=1\columnwidth]{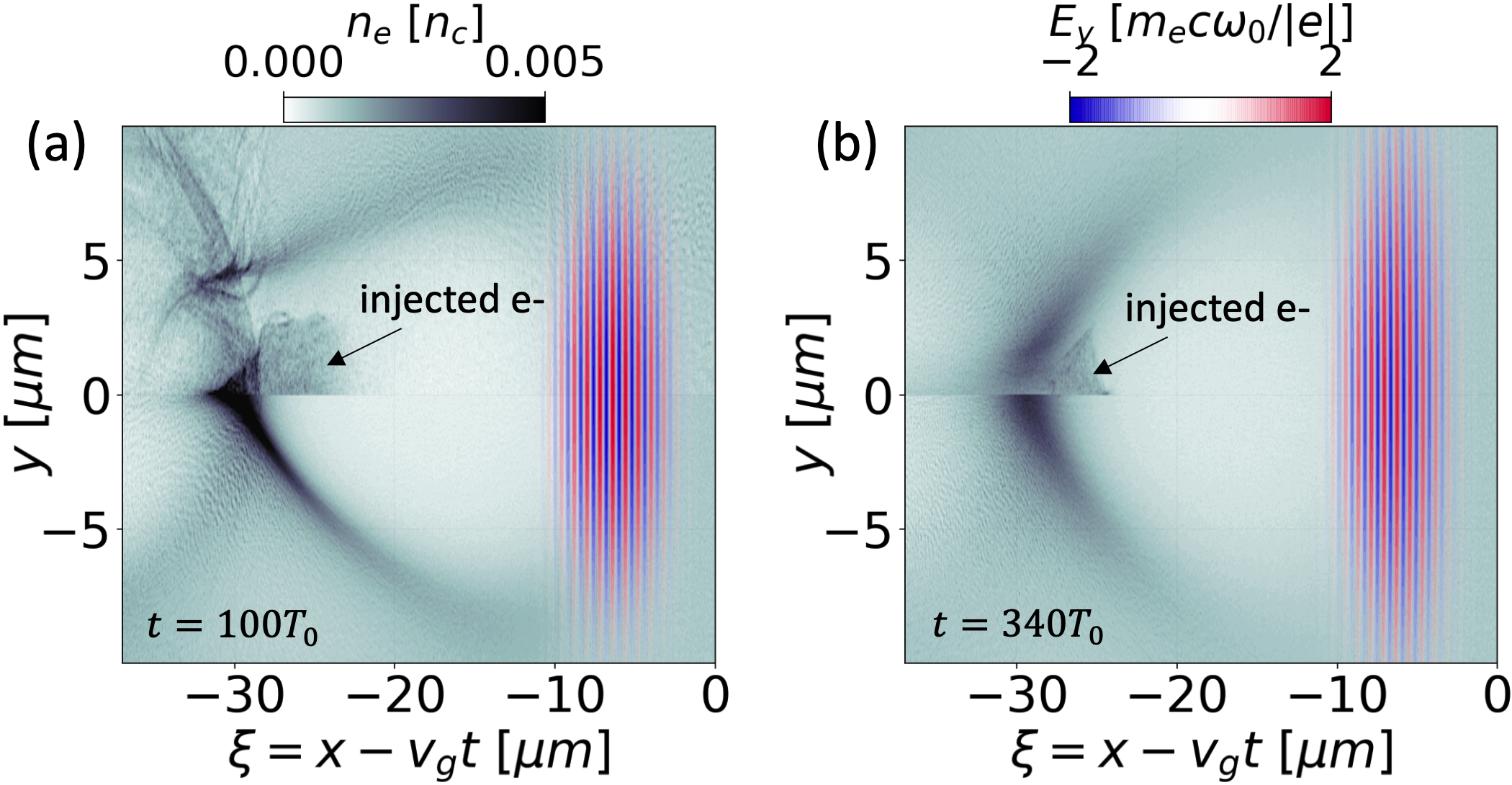}
\caption{2D PIC simulation results. The driving and colliding laser pulse intensities are $a_0=2$ and $a_1=0.5$, respectively. Both pulses have $w_0=\SI{8}{\micro\metre}$ waist radius and $\tau_0=\SI{25}{fs}$ duration. (a) Snapshot of the electron plasma density $n_e$ and the laser electric field $E_y$ at the time $t=100\,T_0$. (b) Same as panel~(a) but at the time $t=340\,T_0$. In panels~(a) and (b), the upper (lower) half panel corresponds to the case with (without) the colliding laser pulse.} \label{fig:den_ey}
\end{figure}
\begin{figure*}[t]
\includegraphics[width=2\columnwidth]{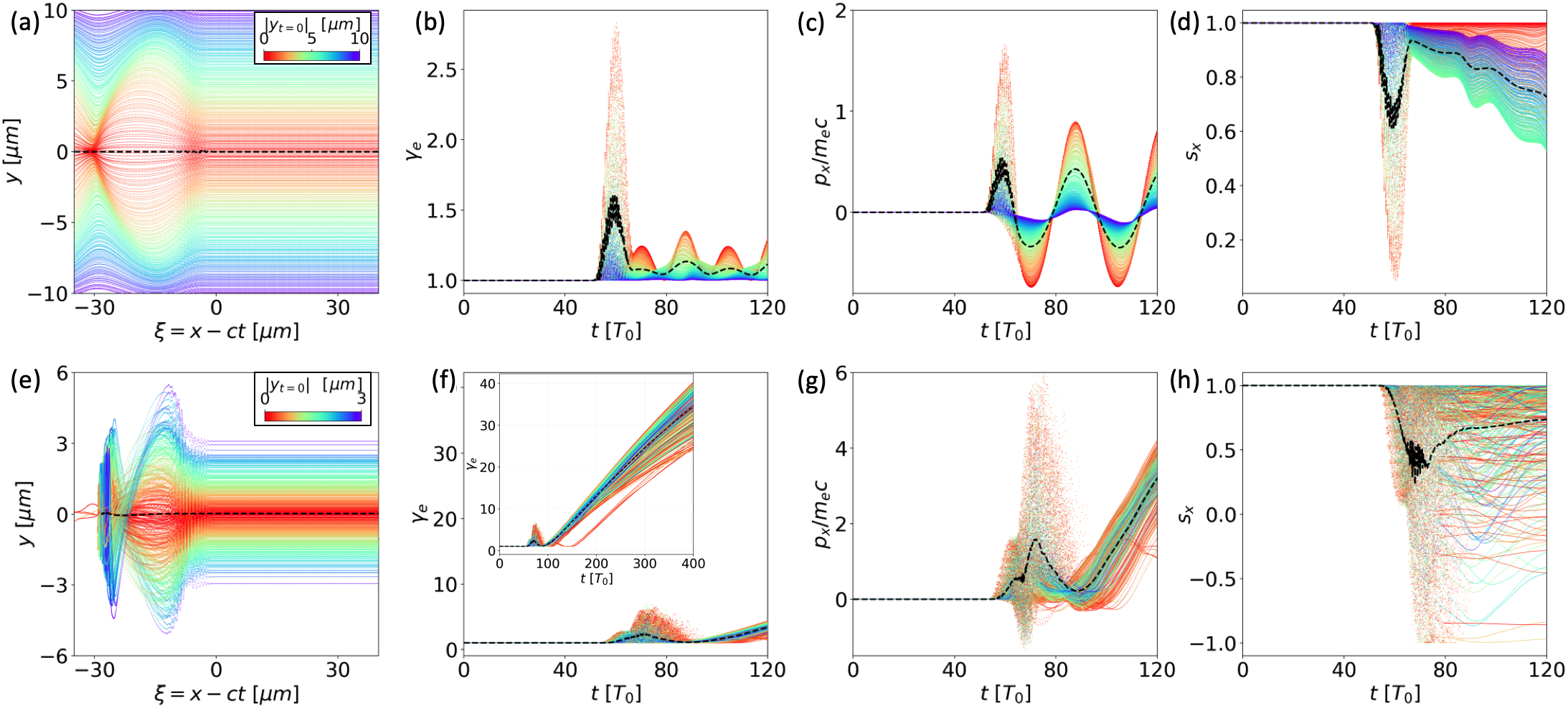}
\caption{Particle tracking results from 2D PIC simulations with the same parameters as those in Fig.~\ref{fig:den_ey}. The rainbow colormap denotes the initial electron's transverse position $|y_{t=0}|$. The black dashed line displays the value obtained by averaging over the displayed trajectories. (a) Electron trajectories in the wake-frame coordinates $(\xi, y)$. (b) The temporal evolution of the electron energy $\gamma_e$. (c) The longitudinal electron momentum $p_x$. (d) The longitudinal spin component $s_x$ of the electron. Panels~(a)--(d) correspond to the case without the colliding laser pulse. Panels~(e)--(h) show the same quantities as (a)--(d) but with the colliding pulse.} \label{fig:tracking}
\end{figure*}

Our 2D simulations were performed using the PIC code EPOCH~\cite{arber2015contemporary} where we implemented the electron spin dynamics. Following Ref.~\cite{wen2019polarized}, we leverage Ehrenfest's theorem~\cite{Ehrenfest1927} to describe the spin of an electron in a quasiclassical state with a vector $\bd{s}$, where $\left| \bd{s} \right| = 1$. The evolution of $\bd{s}$ is determined by the Thomas-Bargmann-Michel-Telegdi (TBMT) equation~\cite{thomas1927kinematics, bargmann1959precession}, and in our EPOCH simulations is implemented via the Boris pusher method~\cite{vieira2011polarized, gong2021retrieving, gong2022deciphering, gong2023electron} (see below and the Appendix for an alternative implementation). Given the relatively low laser pulse intensities considered here, radiation reaction effects~\cite{Tamburini2010, Tamburini2011} as well as other spin effects such as the Stern-Gerlach force~\cite{Wen2016, WenPRA17} and the Sokolov-Ternov effect~\cite{Sokolov1967, thomas2020scaling} are negligible. In simulations, an underdense plasma is irradiated by a relativistic-intensity driving laser pulse and a subrelativistic-intensity colliding laser pulse. The driving pulse is linearly polarized along the $y$ axis, incoming from the left boundary of the computational box, and has a Gaussian transverse and longitudinal profile with $w_0 = \SI{8}{\micro\metre}$ waist radius, $I_0 = 8.7 \times 10^{18}\text{ W/cm}^2$ peak intensity, and $\tau_0 = \SI{25}{fs}$ duration full width at half maximum (FWHM) of the intensity. Its wavelength is $\lambda_0 =\SI{0.8}{\micro\metre}$ and the corresponding normalized field amplitude is $a_0 \approx 0.85 \lambda_0[\SI{}{\micro\metre}] \sqrt{I_0 [10^{18}\text{ W/cm}^2]} \approx 2$. The colliding pulse has the same parameters as those of the driving pulse except for the peak intensity which is $I_1 = 5.4 \times 10^{17} \text{ W/cm}^2$ corresponding to a normalized field amplitude $a_1 \approx 0.5$. 

The computational box size is $120\lambda_0(x) \times 70\lambda_0(y)$ and is uniformly divided in cells with a size of $\lambda_0/20 (x) \times \lambda_0/20 (y)$. The pre-polarized plasma has a plateau electron density profile $n_e(x) = n_{e,0} = 10^{18}\text{ cm}^{-3}$ for $x > x_1$ and a linear up-ramp density profile 
for $x_0 < x < x_1$, where $x_0 = 0$ and $x_1 = 20 \, \lambda_0$ [see Fig.~\ref{fig:schematic}(a)]. In the simulation 32 particles-per-cell are used for electrons, and ions are treated as immobile. The focus of both the driving and the colliding laser pulse is set to be located at $x = 35 \, \lambda_0$ in vacuum. The simulation box moves at the speed of light $c$, and open boundary conditions are adopted for both fields and particles. The group velocity of the driving laser pulse propagating inside the underdense plasma is $v_g = c\sqrt{1-\omega_{pe}^2/(\tilde{\gamma}\omega_0^2)}$, where $\omega_{pe} = \sqrt{4 \pi n_{e} e^2/m_e}$ is the plasma frequency, $\omega_0 = 2\pi c/\lambda_0$ the laser angular frequency, and $\tilde{\gamma} \approx \sqrt{1 + a_0^2/2}$ the cycle-averaged Lorentz factor of the plasma electrons. Here $m_e$ and $e$ are the electron mass and charge, respectively. 

In order to illustrate the effect of the colliding laser pulse on the electron dynamics, we compare the 2D PIC simulation results obtained with and without the colliding pulse. As displayed in Fig.~\ref{fig:den_ey}, a plasma cavity with length $c/\omega_{pe} \approx \SI{30}{\micro\metre}$ is sustained behind the driving pulse. In the presence of the colliding pulse an electron bunch is stably injected at the rear of the cavity resulting in approximately $15$~MeV energy gain over \SI{200}{\micro\metre} propagation distance, whereas essentially no electron injection is observed without the colliding pulse (see Fig.~\ref{fig:den_ey}).
The corresponding particle tracking results from 2D PIC simulations are displayed in Fig.~\ref{fig:tracking}. For the case without colliding pulse, the background plasma electrons merely experience the smooth oscillation excited by the driving laser ponderomotive force, and no background electrons are injected into the wakefield cavity [Fig~\ref{fig:tracking}(a)-(d)]. These electrons do not have a net energy gain, and their growing depolarization over time is attributed to the spin precession induced by the magnetic field while traversing the plasma cavity 
[see Fig.~\ref{fig:tracking}(d)]. By contrast, with colliding pulse a fraction of the electrons originating from the central region is injected into the first wakefield cavity and subsequently undergoes acceleration [Fig~\ref{fig:tracking}(e)-(h)]. Electron injection occurs due to the electrons' residual longitudinal momentum $p_x>0$ after interacting with the colliding laser fields [see Fig.~\ref{fig:tracking}(g) and below]. The electron beam spin polarization is primarily determined by the transient chaotic dynamics induced during the driving- and colliding-pulse interaction, while it is almost unchanged during the acceleration phase [see Fig.~\ref{fig:tracking}(h)]. In contrast with the spin dynamics observed in the down-ramp injection scheme~\cite{wen2019polarized}, in CPI no strong correlation between the accelerated electrons' longitudinal spin polarization loss $1-s_x$ and the electrons' initial transverse coordinate $y$ is observed [see Fig.~\ref{fig:tracking}(h) and section~\ref{section IV}].

\section{\label{section III} Theoretical analysis}

In the following we employ a two-stage model to characterize the electron dynamics and elucidate the injection process.

\subsection{Electron collisionless heating in colliding pulses}

It is known that in vacuum an electron initially at rest remains at rest after interacting with a  laser pulse, if the pulse can be approximated as a plane wave. Thus, in a one-dimensional model where plasma fields are small compared to the laser fields, the electron longitudinal residual momentum $\delta p_x$ mainly stems from the interaction with the fields of the two colliding laser pulses. The residual momentum $\delta p_x$ is of critical importance in determining the electron injection into the forward moving plasma cavity. If the plane-wave fields are derived from a vector potential expressed as $\mathbf{A}_{0,1}$, the corresponding electric and magnetic fields are $\mathbf{E}_{0,1} = -\partial \mathbf{A}_{0,1}/\partial c t$ and $\mathbf{B}_{0,1} = \nabla \times \mathbf{A}_{0,1}$, where the subscript $0$ ($1$) denotes the driving (colliding) laser pulse. By considering for simplicity the vector potential of monochromatic plane waves $\mathbf{A}_0 = a_0 (m_e c^2/|e|) \sin \phi \,\hat{e}_y$ and $\mathbf{A}_1 = a_1 (m_e c^2/|e|) \sin(\phi + 2 k_0 x +\phi_1) \,\hat{e}_y$ with $\phi = \omega_0 t - k_0 x$ being the light front time, $\phi_1$ the initial phase, $\hat{e}_y$ the unit vector along $y$ direction, and $k_0 = \omega_0 /c$ the wavenumber, the electron dynamics inside the two colliding laser pulse fields is determined by 
\begin{align}
\frac{d p_y}{d\phi} = & \frac{|e|}{c}\frac{d }{d\phi}(A_{0,y}+A_{1,y})  \label{eq:px_py_colliding_p} \\
\frac{d p_x}{d\phi} = & -\frac{|e|}{\omega_0}\frac{p_y}{p_{-}}(B_{0,z}+B_{1,z}) \label{eq:px_py_colliding_l}
\end{align}
where $p_{-}\equiv \gamma_e m_e c - p_x$, $d\phi/dt= p_{-} \omega_0 / \gamma_e m_e c$, $B_{0,z} = -a_0 (m_e \omega_0 c/|e|)
\cos\phi $, and $B_{1,z} = a_1 (m_e \omega_0 c/|e|) \cos(\phi + 2 k_0 x + \phi_1)$. From Eq.~\eqref{eq:px_py_colliding_p} one immediately derives an integral of motion for the transverse momentum $p_y = |e| (A_{0,y} + A_{1,y})/c$, such that Eq.~\eqref{eq:px_py_colliding_l} becomes
\begin{align} 
\frac{d p_x}{d\phi} = & \frac{m_e^2 c^2}{p_{-}}[ a_0^2 \cos\phi \sin\phi + a_0a_1 \sin(2 k_0 x + \phi_1) \nonumber \\
& - a_1^2 \cos(\phi+2k_0 x + \phi_1) \sin(\phi + 2 k_0 x + \phi_1)]. \label{eq:px_py_colliding_l2}
\end{align}
The terms containing $2 k_0 x$ in Eq.~\eqref{eq:px_py_colliding_l2} hint at a strong dependence on initial conditions. In fact, previous studies already showed that the resulting dynamics is chaotic~\cite{bauerPRL95}, and that plasma heating due to stochastic acceleration can occur inside the counterpropagating laser pulses~\cite{sheng2002stochastic}.

In order to obtain the dependence of the residual momentum $\delta p_x$ and spin depolarization $\delta s_x$ on the laser parameters, we therefore resort to test particle simulations~\cite{gong2016radiation, gong2019_PRAB, gong2019radiation}. In our test particle simulations, the two laser pulses are modeled as plane waves with electric field
\begin{align} \label{eq:test_laser_field}
\frac{|e| E_{y,0}}{m_e \omega_0 c} = & a_0 \exp\left\{-\left(\frac{\phi-\phi_0}{\omega_0 \tau_0 / \sqrt{2\ln 2}}\right)^2\right\} \cos (\phi), \\
\frac{|e| E_{y,1}}{m_e \omega_0 c} = & a_1 \exp\left\{-\left(\frac{\phi + 2 k_0 x - \phi_1}{\omega_0 \tau_0/\sqrt{2\ln 2}}\right)^2 \right\} \cos(\phi + 2 k_0 x),
\end{align}
and magnetic field $B_{0,z} = E_{0,y}$ and $B_{1,z} = -E_{1,y}$. The laser 
wavelength and period are denoted as $\lambda_0 = \SI{0.8}{\micro\metre}$ and $T_0 = \lambda_0/c \approx \SI{2.67}{fs}$, respectively, while $\tau_0$ is the FWHM of the intensity. Here $\phi_0 = 0$ and $\phi_1 = 100 \pi$ determine the initial position of the peak of the driving and colliding laser pulses, which correspond to $0$ and $50\lambda_0$, respectively. The initially at rest and uniformly distributed electrons are located in the region $20 \lambda_0 \leqslant x \leqslant 30 \lambda_0$. In calculating the electron momentum $\bd{p}$ and spin $\bd{s}$ an explicit Boris pusher method is utilized, where the timestep is $\Delta t = 5 \times 10^{-4}\,T_0$. This timestep fulfills the stringent temporal criteria of electron acceleration~\cite{arefiev2015temporal}.
\begin{figure}[tb]
\includegraphics[width=0.75\columnwidth]{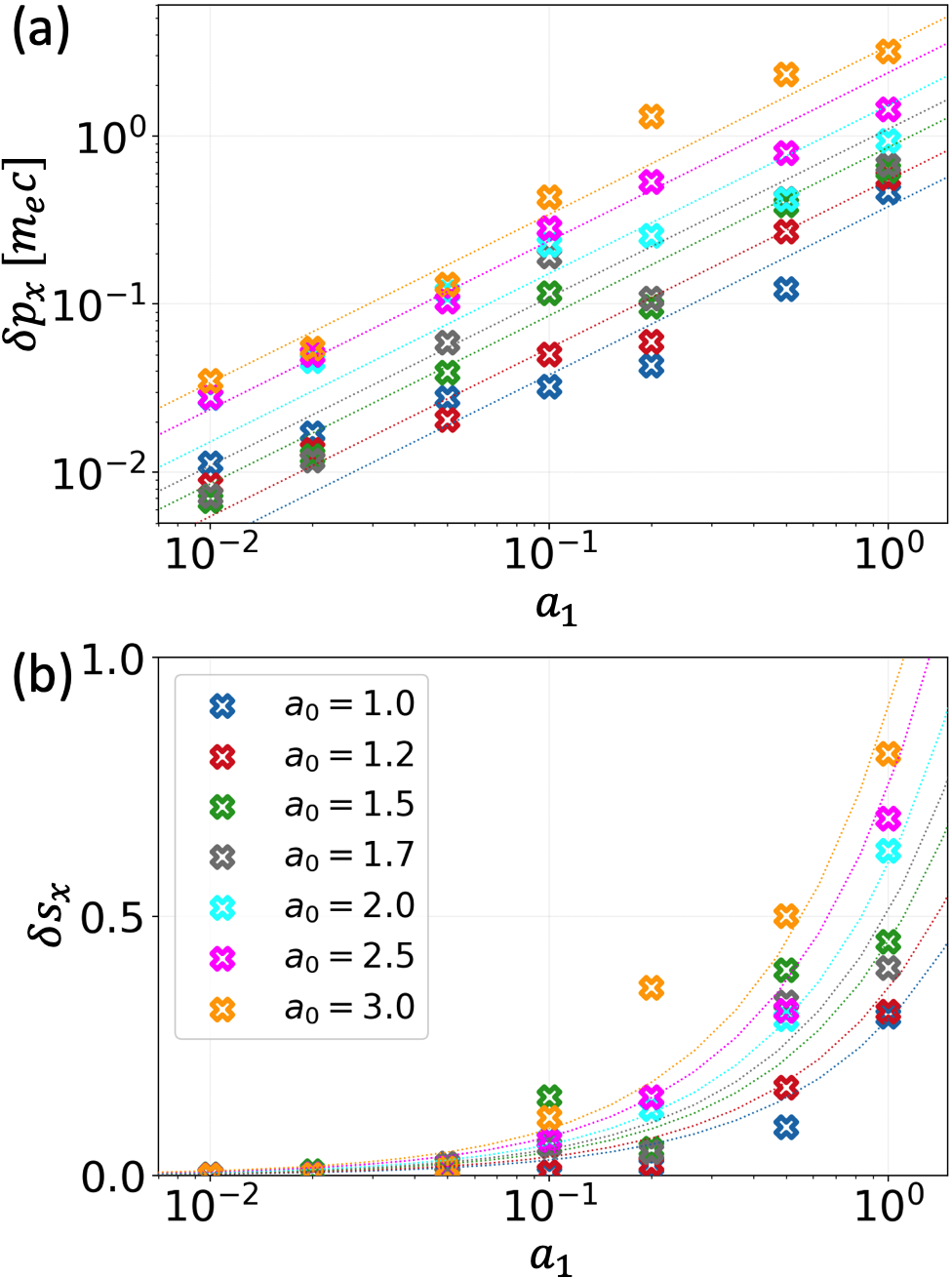}
\caption{Test particle simulation results. Each color corresponds to a different driving laser pulse amplitude $a_0$ and \bl{the horizontal axis presents the colliding laser pulse amplitude $a_1$}. (a) Residual longitudinal momentum $\delta p_x$ after the collision of the two plane-wave pulses. (b) Spin polarization loss $\delta s_x \equiv 1 - s_x$. In both panels, dashed lines display the prediction obtained by numerical fitting the simulation data as $\delta p_x = 0.29 a_0^2 a_1 m_e c$ and $\delta s_x = 0.25 a_0a_1$.} \label{fig:test_pxsx}
\end{figure}
\begin{table}[bt]
\caption{The parameters of the scaling $\delta p_x \approx \kappa_p a_0^{n_0} a_1^{n_1} m_e c$ calculated by numerical fitting the results of test particle simulations.}
\label{tab:px_scaling}
\tabcolsep=0.25cm
\begin{tabular}{c | c c c c c c c}
\hline
\hline
$\tau_0\ [\SI{}{fs}]$ & 6.2 & 12.6 & 18.8 & 25.0 & 31.4 & 37.7 & 43.9 \\
  \hline  
$n_0$ & 0.75 & 1.25 & 2.0 & 2.0 & 2.0 & 3.0 & 3.25 \\
  \hline
$n_1$ & 0.75 & 0.75 & 1.0 & 1.0 & 1.0 & 1.0 & 1.0\\
  \hline
$\kappa_p$ & 0.30 & 0.26 & 0.27 & 0.29 & 0.32 & 0.27 &0.28 \\
  \hline
 \hline
  \end{tabular}
\end{table}
\begin{table}[bt]
\caption{The parameters of the scaling $\delta s_x \approx \kappa_s a_0^{m_0}a_1^{m_1}$ calculated by numerical fitting the results of test particle simulations.} \label{tab:sx_scaling}
\tabcolsep=0.25cm
\begin{tabular}{c | c c c c c c c}
\hline
\hline
$\tau_0\ [\SI{}{fs}]$ & 6.2 & 12.6 & 18.8 & 25.0 & 31.4 & 37.7 & 43.9 \\
  \hline
$m_0$ & 1.0 & 1.0 & 1.0 & 1.0 & 1.0 & 1.5 & 1.5 \\
  \hline
$m_1$ & 1.0 & 1.0 & 1.0 & 1.0 & 1.0 & 1.0 & 1.0\\
  \hline
$\kappa_s$ & 0.10 & 0.17 & 0.19 & 0.25 & 0.27 & 0.30 & 0.36\\
  \hline
 \hline
  \end{tabular}
\end{table}

For $\tau_0=\SI{25}{fs}$, the test particle simulation results for the residual longitudinal momentum $\delta p_x$ and spin variation $\delta s_x$ are shown in Fig.~\ref{fig:test_pxsx}. Both $\delta p_x$ and $\delta s_x$ are calculated by averaging over the forward-moving electrons after they are separated from the two colliding pulses. By numerically fitting the results over the range $1\leqslant a_0 \leqslant 3$ and $10^{-2}\leqslant a_1\leqslant 1$, we obtain the scaling $\delta p_x \approx 0.29 a_0^2a_1 m_e c$ and $\delta s_x \approx 0.25 a_0 a_1$. As shown in Fig.~\ref{fig:test_pxsx} the curves obtained from the above simple scaling model fairly agree with the test particle simulation results.

It is worth emphasizing that, in general, stochastic heating and consequently $\delta p_x$ and $\delta s_x$, is also expected to depend on the laser pulse duration $\tau_0$. To examine the impact of $\tau_0$, for each $\tau_0$ in the range $\SI{6.2}{fs} \leqslant \tau_0 \leqslant \SI{43.9}{fs}$ we assume a scaling of the form $\delta p_x = \kappa_p a_0^{n_0} a_1^{n_1} m_e c$ and $\delta s_x = \kappa_s a_0^{m_0} a_1^{m_1}$, where $\kappa_{p,s}$, $n_{0,1}$, and $m_{0,1}$ are constants obtained by numerically fitting the residual longitudinal momentum and electron spin. Tables~\ref{tab:px_scaling} and \ref{tab:sx_scaling} report the obtained coefficients. Table~\ref{tab:px_scaling} highlights a pronounced dependence of the exponent $n_0$ on the laser pulse duration, which originates from an increased stochastic heating and longitudinal momentum gain of electrons for longer duration laser pulses.
Given the relative simplicity of the obtained scaling, this is employed for quantitative predictions of the electron injection threshold and of the final beam polarization, which are validated against PIC simulations (see below).

\subsection{Hamiltonian analysis of electron trapping}

The second stage of electron injection corresponds to electron trapping into the subluminal wakefield, which is investigated through Hamiltonian analysis.

In laser-driven wakefield acceleration, electrons gain energy from the longitudinal electric field of the Langmuir wave excited by the ponderomotive force of the laser pulse. This is modeled, for simplicity, by considering the one-dimensional dynamics of electrons in the moving frame of the first cavity in the wake of the laser pulse. The drifting velocity of the plasma cavity $v_d$ equals the group velocity of the laser pulse inside the underdense plasma $v_g$, i.e., $v_d = c\sqrt{1 - \omega_{pe}^2/(\tilde{\gamma} \omega_0^2)}$. In the cavity-frame coordinate $\xi \equiv x - v_dt$, the longitudinal electric field $E_x(\xi)$ depends only on $\xi$. The electron dynamics is characterized by the equations~\cite{esareyPoP95}
\begin{align}
\frac{d p_x}{dt} = & -|e|E_x(\xi), \label{eq:dp}  \\
\frac{d \xi}{dt} = & \frac{p_x}{m_e\sqrt{1+(p_x/m_ec)^2}}-v_d , \label{eq:dxi}
\end{align}
where $v_d$ is independent of time. The electric potential of the longitudinal wakefield can be derived as $\varphi(\xi)=-\int E_x(\xi) d\xi$ such that $E_x(\xi)=-\partial \varphi(\xi)/\partial \xi$. This allows us to determine the electron motion in the moving wakefield from the conserved Hamiltonian:
\begin{equation}
\mathcal{H}(\xi,p_x)=-|e|\varphi(\xi) + c\sqrt{m_e^2c^2+p_x^2}-v_dp_x \label{eq:Hami}
\end{equation}
Indeed, Eqs.~\eqref{eq:dp}-\eqref{eq:dxi} can be obtained from Hamilton's equations $d p_x/dt=-\partial \mathcal{H}(\xi,p_x)/\partial \xi$ and $d \xi/dt = \partial \mathcal{H}(\xi,p_x)/\partial p_x$. 
By setting $dp_x/dt=0$ and $d\xi/dt=0$, we find a fixed point $(\xi^*, p_x^*)$ in the ($\xi, p_x$) phase space where $\xi^*$ satisfies the condition $E(\xi^*)=0$ and $p_x^* = (v_dm_e)/\sqrt{1 - v_d^2/c^2}$. The fixed point corresponds to a scenario in which an electron with velocity $v_x = v_d$ is co-moving with the wakefield and does not exchange energy with the longitudinal electric field. 

\begin{figure}[tb]
\includegraphics[width=1\columnwidth]{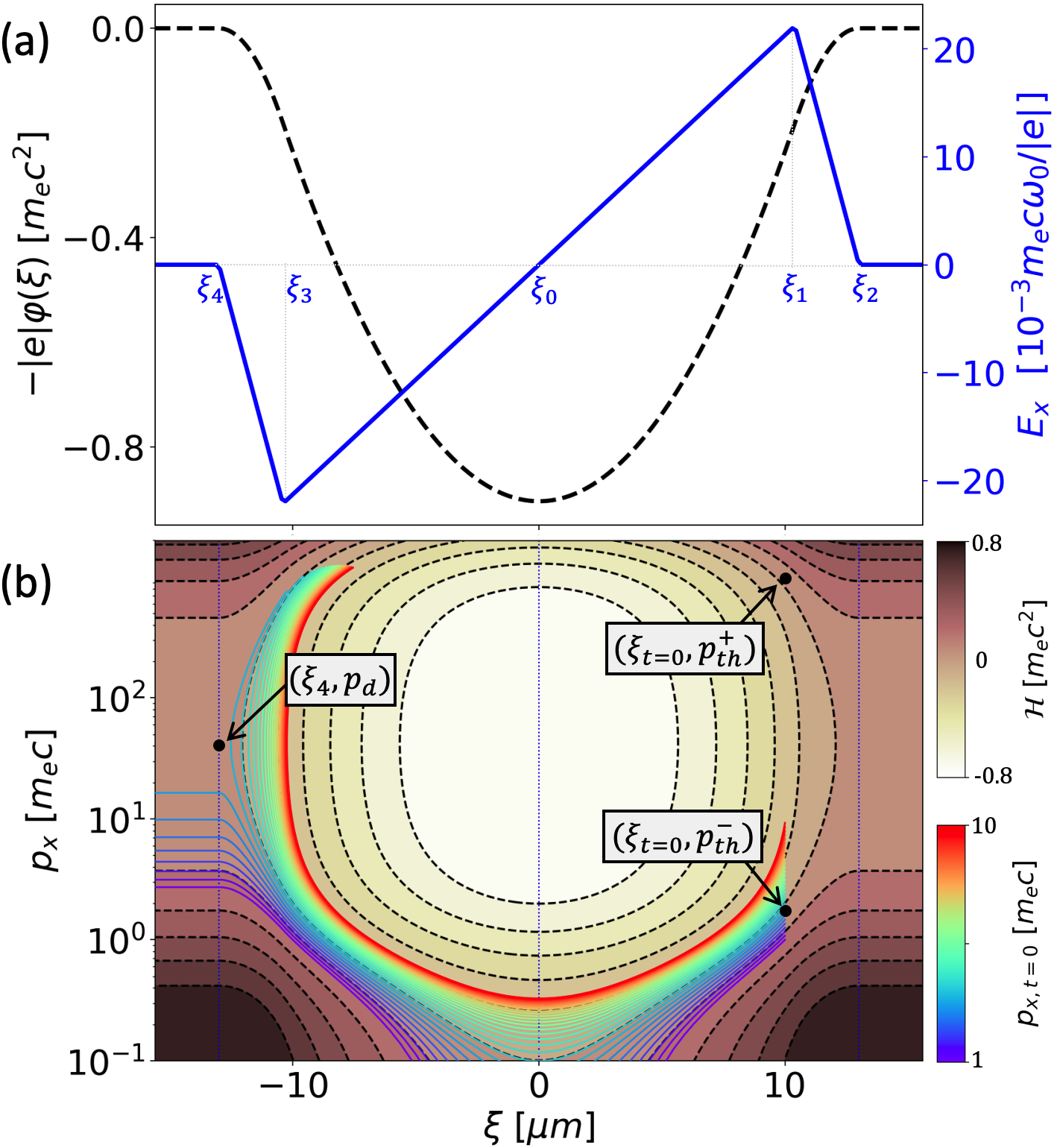}
\caption{Illustration of the the Hamiltonian model. (a) The electron potential energy $-|e|\varphi$ (black dashed line) and longitudinal electric field $E_x$ (blue solid line) as a function of the wake-frame coordinate $\xi$. (b) The value of the Hamiltonian $\mathcal{H}(\xi, p_x)$ in Eq.~\eqref{eq:Hami} in units of electron rest energy $m_e c^2$ (brown color map) and its contour levels (black dashed lines). The rainbow color lines display the evolution in the $(\xi, p_x)$ phase space of the electrons initially located at $\xi = \SI{10}{\micro\metre}$. Panels~(a) and (b) share the same horizontal axis.} \label{fig:hami_th}
\end{figure}

We are interested in the electron dynamics inside the first cavity of the laser-driven wake. The corresponding longitudinal electric field can be approximated as [see Fig.~\ref{fig:hami_th}(a)]
\begin{eqnarray}
E_x(\xi) =
\left\{
	\begin{array}{ll}
	    0,  & \mbox{if } \xi \leqslant \xi_4 \\
		-E_0\frac{\xi-\xi_4}{\xi_3-\xi_4}, & \mbox{if } \xi_4< \xi \leqslant \xi_3 \\
	    -E_0\frac{\xi_0-\xi}{\xi_0-\xi_3}, & \mbox{if } \xi_3< \xi \leqslant \xi_0 \\
		
		E_0\frac{\xi-\xi_0}{\xi_1-\xi_0}, & \mbox{if } \xi_0< \xi \leqslant \xi_1 \\
	    E_0\frac{\xi_2-\xi}{\xi_2-\xi_1}, & \mbox{if } \xi_1< \xi \leqslant \xi_2 \\
	    0,  & \mbox{if } \xi_2 < \xi
	\end{array} \label{eq:modelEx}
\right.
\end{eqnarray}
where $E_0$ is the peak value of $|E_x(\xi)|$ and is reached at $\xi_{1,3}$, while $\xi_{2,4}$ denote the boundaries of the cavity. As shown in Fig.~\ref{fig:hami_th}(a), $\xi_4=-\xi_2$ and $\xi_3=-\xi_1$ as a result of the symmetry of the field with respect to $\xi_0=0$. Accordingly, the potential is calculated through $\varphi(\xi)=-\int_{-\infty}^\xi E(\xi) d\xi$ with $\varphi(-\infty)=0$, which gives
\begin{eqnarray}
\varphi(\xi) =
\left\{
	\begin{array}{ll}
	    0,  & \mbox{if } \xi \leqslant \xi_4 \\
		E_0\frac{(\xi-\xi_4)^2}{2(\xi_3-\xi_4)}, & \mbox{if } \xi_4< \xi \leqslant \xi_3 \\
	    E_0\frac{\xi_0-\xi_4}{2}-E_0\frac{(\xi_0-\xi)^2}{2(\xi_0-\xi_3)}, & \mbox{if } \xi_3< \xi \leqslant \xi_0 \\
		
		E_0\frac{\xi_2-\xi_0}{2}-E_0\frac{(\xi-\xi_0)^2}{2(\xi_1-\xi_0)}, & \mbox{if } \xi_0< \xi \leqslant \xi_1 \\
	    E_0\frac{(\xi-\xi_2)^2}{2(\xi_2-\xi_1)}, & \mbox{if } \xi_1< \xi \leqslant \xi_2 \\
		0 & \mbox{if } \xi_2 < \xi.
	\end{array} \label{eq:modelphi}
\right.
\end{eqnarray}
For definiteness and without loss of generality, we consider the following parameters $v_d/c \approx 0.9997$, $\xi_0 = 0$, $\xi_1 = -\xi_3 = \SI{10.4}{\micro\metre}$, $\xi_2 = -\xi_4 = \SI{13.0}{\micro\metre}$, and $E_0 = 0.022 m_e c \omega_0/|e| \approx 96\,\mathrm{GV/m}$. The above parameters are similar to those identified in the LWFA experiments where GeV electron beams are produced from a centimeter-scale underdense plasma~\cite{leemans2006gev}. The corresponding minimum electron potential energy is $-|e|\varphi(\xi_0)/(m_e c^2)\approx 0.9$.

For the above-mentioned parameters, Fig.~\ref{fig:hami_th}(a) displays the profiles of the electric field $E_x(\xi)$ and potential $\varphi(\xi)$ obtained from Eq.~(\ref{eq:modelEx}) and Eq.~(\ref{eq:modelphi}), respectively. Figure~\ref{fig:hami_th}(b) displays the corresponding values of the Hamiltonian $\mathcal{H}(\xi, p_x)$ as well as the phase space evolution of a group of electrons initially located at $\xi = 
\SI{10}{\micro\metre}$ with momentum $1 \, m_e c \leqslant p_x|_{t=0} \leqslant 10 \, m_e c$. Their evolution in $(\xi, p_x)$ clearly shows that there exists a longitudinal momentum threshold $p_{th}$ for the occurrence of electron trapping in the wake. This allows us to determine whether an energized electron gets trapped by the wakefield $E_x(\xi)$ or slides away from the potential cavity [see Fig.~\ref{fig:hami_th}(b)]. The electrons with $p_x|_{t=0}<p_{th}$ are not sufficiently fast to be trapped by the forward-moving wake. Thus, they slide away from the wake cavity and are not injected. These electrons are termed \textit{untrapped electrons}. By contrast, the electrons with $p_x|_{t=0}>p_{th}$ are trapped by the potential well $\varphi(\xi)$ and subsequently efficiently accelerated to an energy of approximately 400~MeV in the region of the cavity where the field $E_x(\xi)$ is negative and therefore accelerating for electrons. These electrons are termed \textit{trapped electrons}. In order to determine the threshold $p_{th}$, we consider the contour of $\mathcal{H}(\xi, p_x)$ between the separatrix point $(\xi_4, p_d)$ and the threshold $(\xi,p_{th})$, which is given by $\mathcal{H}(\xi, p_{th}) = \mathcal{H}(\xi_4, p_d)$. This gives
\begin{equation} \label{eq:trap_p_th_1}
(1-\beta_d^2)\left( \frac{p_{th}}{m_e c}\right)^2 - 2\beta_d \mathcal{A}\left(\frac{p_{th}}{m_ec}\right)+1 - \mathcal{A}^2 = 0,
\end{equation}
where $\beta_d = v_d/c$, $\mathcal{A} = \frac{|e|\varphi(\xi)}{m_e c^2} + \frac{1}{\gamma_d}$, $\gamma_d = 1/\sqrt{1-\beta_d^2}$, and $p_d = \gamma_d v_d m_e c$. The two solutions $p_{th}^{\pm}$ of Eq.~\eqref{eq:trap_p_th_1} are:
\begin{equation}
\frac{p_{th}^{\pm}}{m_ec} = \frac{\beta_d\mathcal{A}\pm\sqrt{\beta_d^2+\mathcal{A}^2-1}}{1-\beta_d^2}. \label{eq:trap_p_th}
\end{equation}
By employing the parameters below Eq.~\eqref{eq:modelphi}, we obtain $|e| \varphi(\xi)/(m_e c^2) \approx -0.24$ at $\xi = \SI{10}{\micro\metre}$ and $\gamma_d \approx 41.6$. Thus, the momentum threshold for electron trapping is $p_{th}^- \approx \, 1.7 \, m_e c$, which agrees well with the numerically calculated electron trajectories in Fig.~\ref{fig:hami_th}(b). The conjugate root $p_{th}^+ \approx 917 \, m_e c$ corresponds to the attainable energy of an electron trapped with momentum near the threshold after it undergoes acceleration in the cavity and returns to $\xi = \SI{10}{\micro\metre}$ [see Fig.~\ref{fig:hami_th}(b)]. In this description, the maximum and the minimum longitudinal momentum are reached at $\xi_0 = 0$. \bl{Note that, while the simplified longitudinal electric field profile in Eq.~\eqref{eq:modelEx} does not precisely match that obtained in PIC simulations [compare Fig.~\ref{fig:hami_th}(a) and Fig.~\ref{fig:hami_sim}(a)], our analysis and model are not sensitive to the exact form of the longitudinal electric field. In fact, the injection criterion $p_{th}^-$ and the maximum attainable energy $\gamma_e^{max} \sim p_{th}^+$ in Eq.~\eqref{eq:trap_p_th} are determined once the potential $\varphi(\xi)$ around the peak of the longitudinal electric field and the drifting velocity $v_d$ are given (see section~\ref{section IV} for details).}

\section{\label{section IV} Model validation}

\begin{figure}[tb]
\includegraphics[width=1\columnwidth]{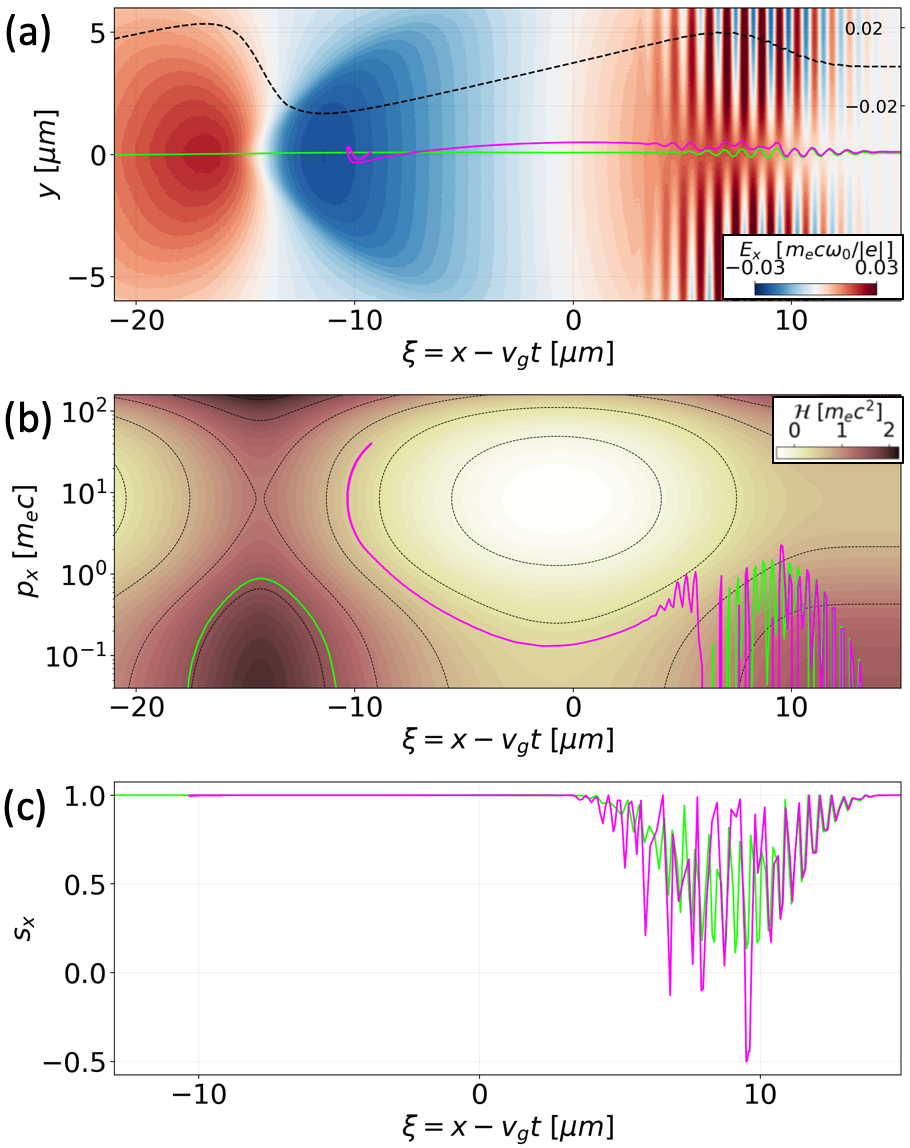}
\caption{Particle tracking results from 2D PIC simulations. The driving- and colliding-laser pulse intensities are $a_0=2$ and $a_1=0.5$, respectively. Both laser pulses have $w_0=\SI{8}{\micro\metre}$ waist radius and $\tau_0=\SI{25}{fs}$ duration. The magenta and green lines correspond to the case with and without the colliding laser pulse, respectively. (a) Electron trajectories in the $(\xi, y)$ space. The blue-red color map displays the longitudinal electric field. The black dashed line plots $E_x$ at $y=0$. (b) Electron trajectories in the $(\xi, p_x)$ space. The brown color map shows the normalized value of the Hamiltonian $\mathcal{H}$ from Eq.~\eqref{eq:Hami}, where the potential $\varphi(\xi)$ is obtained from the $E_x$ at $y=0$ of the simulation [see the black dashed line in panel (a)]. (c) Evolution of the longitudinal spin $s_x$.} \label{fig:hami_sim}
\end{figure}
\begin{figure*}[tb]
\includegraphics[width=1.6\columnwidth]{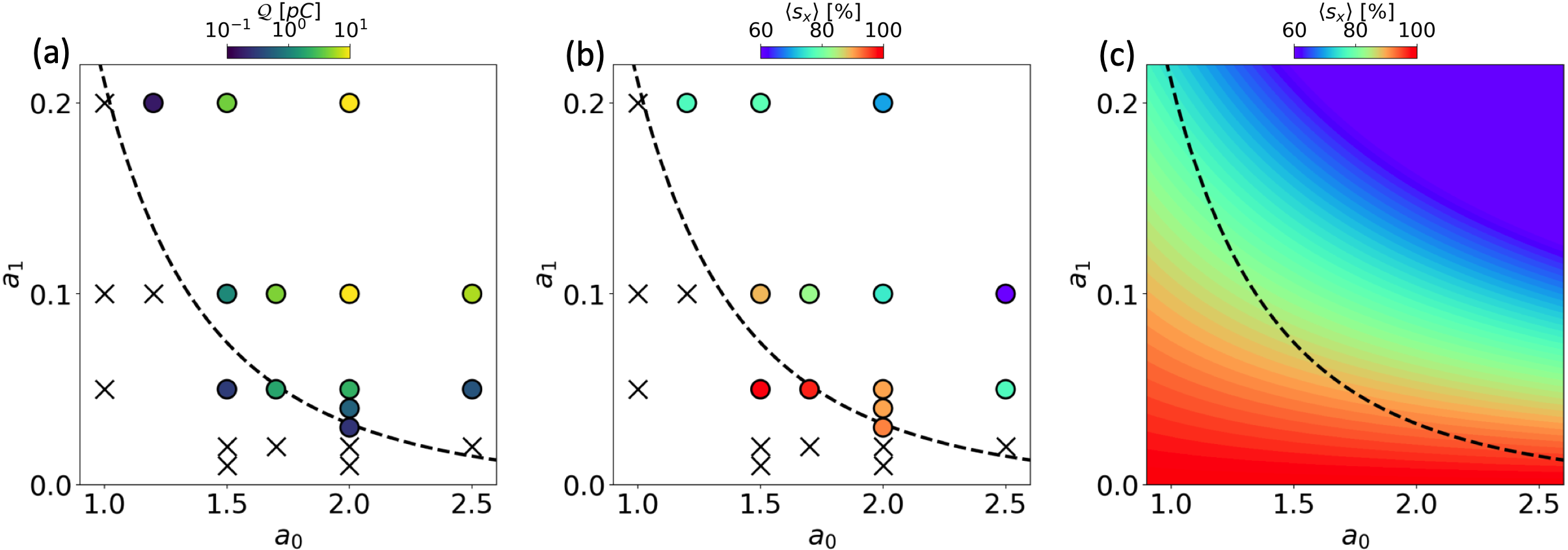}
\caption{Parameter scans over the driving $a_0$ and colliding $a_1$ laser pulse normalized amplitude performed with the spectral quasi-3D PIC code FBPIC. (a) Injected electron charge $\mathcal{Q}$. (b) electron beam average spin polarization $\left\langle s_x \right\rangle$. The cross marks in panels~(a) and (b) denote the cases in which no significant electron injection was observed. The black dashed line in panels (a)-(c) plots the injection threshold according to Eq.~\eqref{eq:injection_criterion}. (c) Average longitudinal spin polarization $\left\langle s_x\right\rangle  = 1 - \kappa_s a_0 a_1$ as predicted from the scaling obtained with the test particle simulations (see Tab.~\ref{tab:sx_scaling}).} \label{fig:injection}
\end{figure*}

In order to validate the model presented in Sec.~\ref{section III} and the injection condition $\delta p_x > p_{th}^{-}$, with $p_{th}^{-}$ defined in Eq.~\eqref{eq:trap_p_th}, we track the evolution of an electron with initial position $y \approx 0$ in the 2D PIC simulations of section~\ref{section II} and investigate its evolution both with and without the colliding pulse. The corresponding results are displayed in Fig.~\ref{fig:hami_sim}, where the magenta and green lines correspond to the case with and without the colliding pulse, respectively. 
In both cases, the electron trajectory in the $(\xi, p_x)$ space shows that its evolution nearly follows the contour of the Hamiltonian~\footnote{Here the potential $\varphi(\xi)$ in the Hamiltonian $\mathcal{H}$ is obtained from the $E_x$ at $y=0$ of the 2D PIC simulation [see the black dashed line in Fig.~\ref{fig:hami_sim}(a)].} after the electron interaction with the pulses ends [see Fig.~\ref{fig:hami_sim}(b)]. In the case without the colliding pulse, the electron trajectory in $(\xi, p_x)$ remains always below the separatrix, the electron is not trapped and readily slides away from the plasma cavity. In the case with the colliding pulse, however, the electron has a residual longitudinal momentum $\delta p_x > 0$. The residual momentum satisfies the injection criterion, namely $\delta p_x > p_{th}^-$, and the electron gets trapped in the cavity [see Fig.~\ref{fig:hami_sim}(b)]. For the traced electron, the longitudinal spin polarization $s_x$ is modulated by the colliding laser fields but returns nearly to its original value after the passage of the laser pulses. Moreover, $s_x$ does not significantly change during the subsequent acceleration stage inside the cavity.

For a homogeneous plasma, one can infer $p_{th}^{\pm}$ in Eq.~\eqref{eq:trap_p_th} with the following estimates: $\beta_d = \sqrt{1 - \mathcal{S}}$, $1/\gamma_d = \sqrt{\mathcal{S}}$, $\mathcal{A} = \rho \Tilde{\varphi} + 1/\gamma_d$, where $\mathcal{S} \equiv n_e/\tilde{\gamma}n_c$, $n_c = m_e \omega_0^2 / 4 \pi e^2$, $\Tilde{\varphi}=|e| \varphi_0 / m_e c^2$, $\varphi_0 = 4 \pi |e| n_e(c/\omega_{pe})^2$, $\tilde{\gamma} = \sqrt{1 + a_0^2/2}$. The coefficient $0 < \rho \lesssim 1$ accounts for the unknown position of the electron inside the cavity when the electron-laser pulses interaction ends [see the magenta line in Fig.\ref{fig:hami_sim}] and for the minimum of the actual potential, which is simply estimated as $\varphi_0$. As it will be clear below, in practice $\rho$ is extracted from quasi-3D PIC simulations. Now, $p_{th}^{-}$ in Eq.~\eqref{eq:trap_p_th} can be recast as
\begin{equation} \label{eq:trap_p_th2} 
\frac{p_{th}^{-}}{m_ec} \approx \frac{\sqrt{1-\mathcal{S}}(\rho\Tilde{\varphi}+\sqrt{\mathcal{S}})-\sqrt{\rho^2\Tilde{\varphi}^2+2\rho\Tilde{\varphi}\sqrt{\mathcal{S}}}}{\mathcal{S}}.
\end{equation}

By combining Eq.~\eqref{eq:trap_p_th2} with the scaling of the residual longitudinal momentum $\delta p_x \approx \kappa_p a_0^{n_0} a_1^{n_1} m_e c$ (see Tab.~\ref{tab:px_scaling}), the injection criterion $\delta p_x > p_{th}^-$ becomes $a_1 > a_1^*$ where
\begin{equation} \label{eq:injection_criterion}
a_1^* = \left[\frac{\sqrt{1-\mathcal{S}}(\rho\Tilde{\varphi}+\sqrt{\mathcal{S}})-\sqrt{\rho^2\Tilde{\varphi}^2+2\rho\Tilde{\varphi}\sqrt{\mathcal{S}}}}{\kappa_p a_0^{n_0}\mathcal{S}}\right]^{1/n_1}.
\end{equation}
\begin{figure*}[tb]
\includegraphics[width=2\columnwidth]{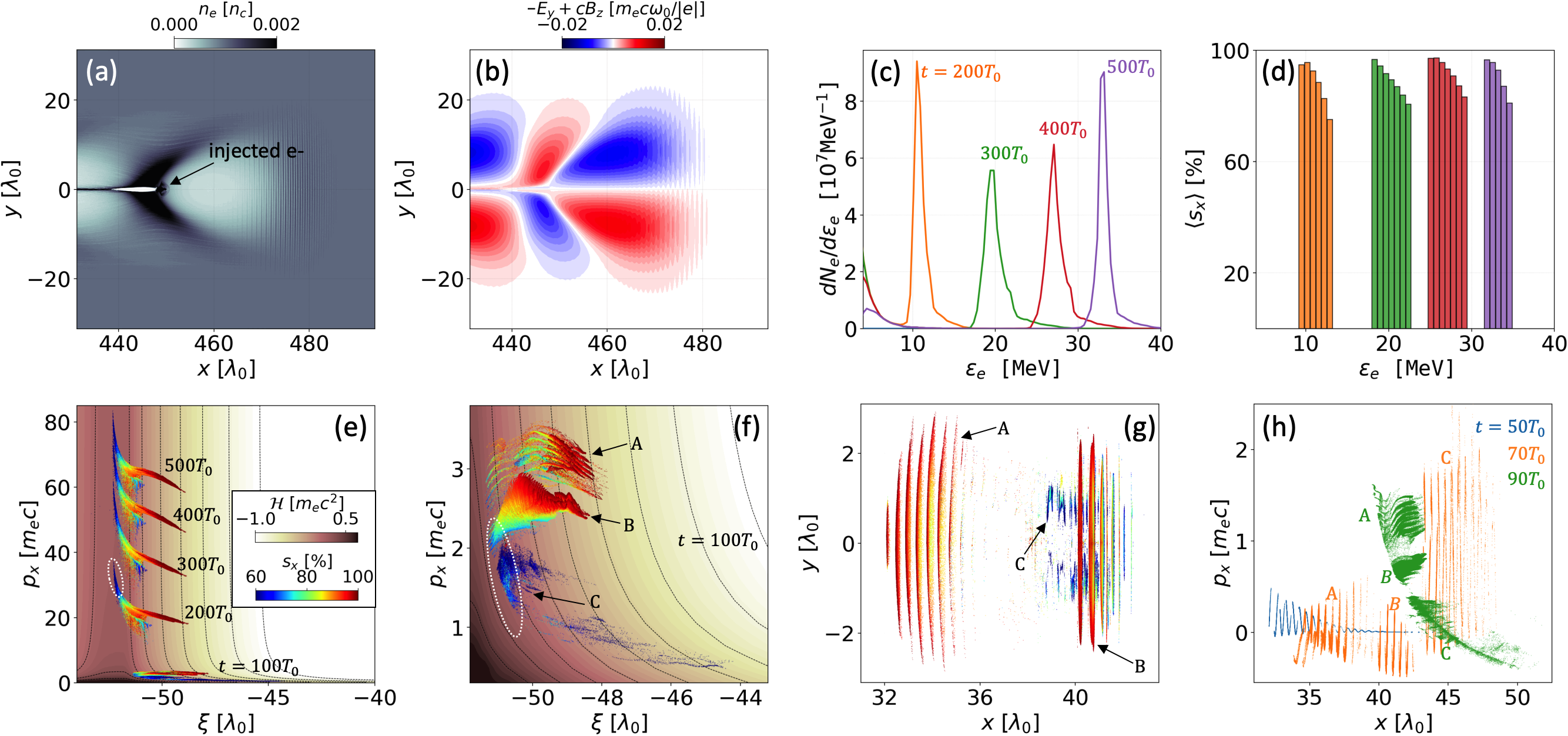}
\caption{FBPIC simulation results with $a_0=2$ and $a_1=0.05$ driving- and colliding-laser pulses, respectively. Snapshot of (a) electron density distribution $n_e$, and (b) transverse focusing force $-E_y+cB_z$ at $t=500\,T_0$. (c) Electron energy spectrum $dN_e / d\varepsilon_e$. (d) Average spin polarization $\left\langle s_x\right\rangle$ as a function of the electron energy $\varepsilon_e$. In panels~(c) and (d) each color refers to a specific time. (e) Evolution of the injected electrons (rainbow color map) in the $(\xi, p_x)$ space and the corresponding Hamiltonian distribution $\mathcal{H}(\xi,p_x)$ (brown color map). (f) Zoom of panel (e) at $t=100\,T_0$ showing the three electron populations labeled A, B, and C. In panels~(e) and (f), the white dashed ellipse marks the electrons near the Hamiltonian separatrix. (g) Initial position in the $(x,y)$ space of the injected electrons that eventually constitute the three populations A, B, C whose evolution is shown in panels~(e) and (f). The rainbow color map in panels~(e)--(g) indicates the spin polarization at the time $t=500\,T_0$. (h) Evolution of injected electron populations in the longitudinal phase space $(x, p_x)$, each color corresponding to a different time, namely, $t=50$, 70, and 90$\,T_0$.} \label{fig:fbpic3d}
\end{figure*}
\begin{figure*}[tb]
\includegraphics[width=1.5\columnwidth]{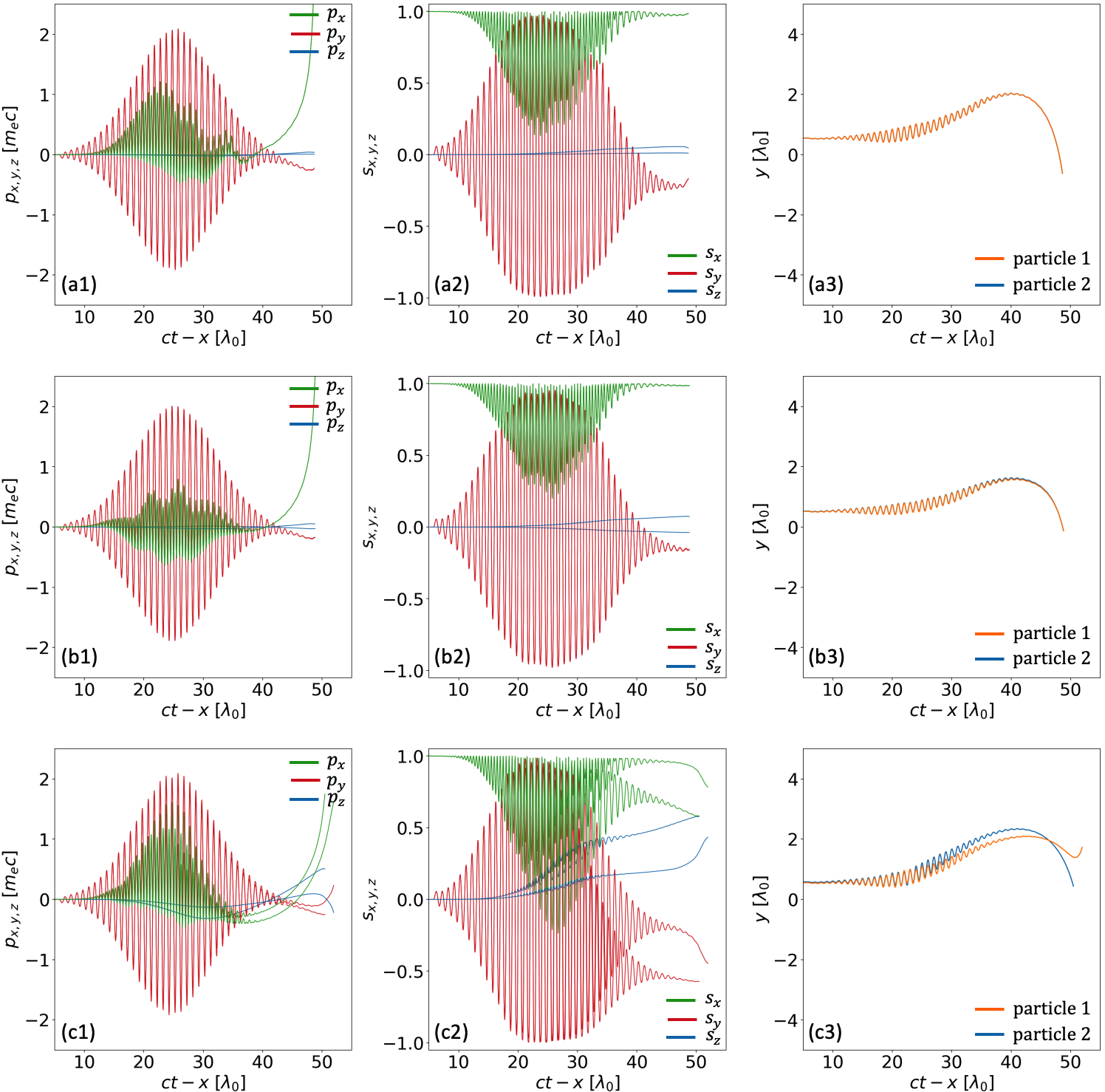}
\caption{FBPIC particle tracking results with $a_0 = 2$ and $a_1 = 0.05$ driving- and colliding-laser pulses, respectively, as a function of $ct-x$, i.e., time evolution is from left to right. (a1) Evolution of the momentum components $p_{x}$ (green), $p_{y}$ (red), $p_{z}$ (blue), and (a2) evolution of the spin components $s_{x}$ (green), $s_{y}$ (red), $s_{z}$ (blue) as well as the (a3) transverse coordinate $y$ of two representative electrons belonging to population~A. Panels~(b1)--(b3) and (c1)--(c3) plot the same quantities as panels~(a1)--(a3) but for population~B and C, respectively.}
\label{fig:s_track}
\end{figure*}
\begin{figure}[tb]
\includegraphics[width=1\columnwidth]{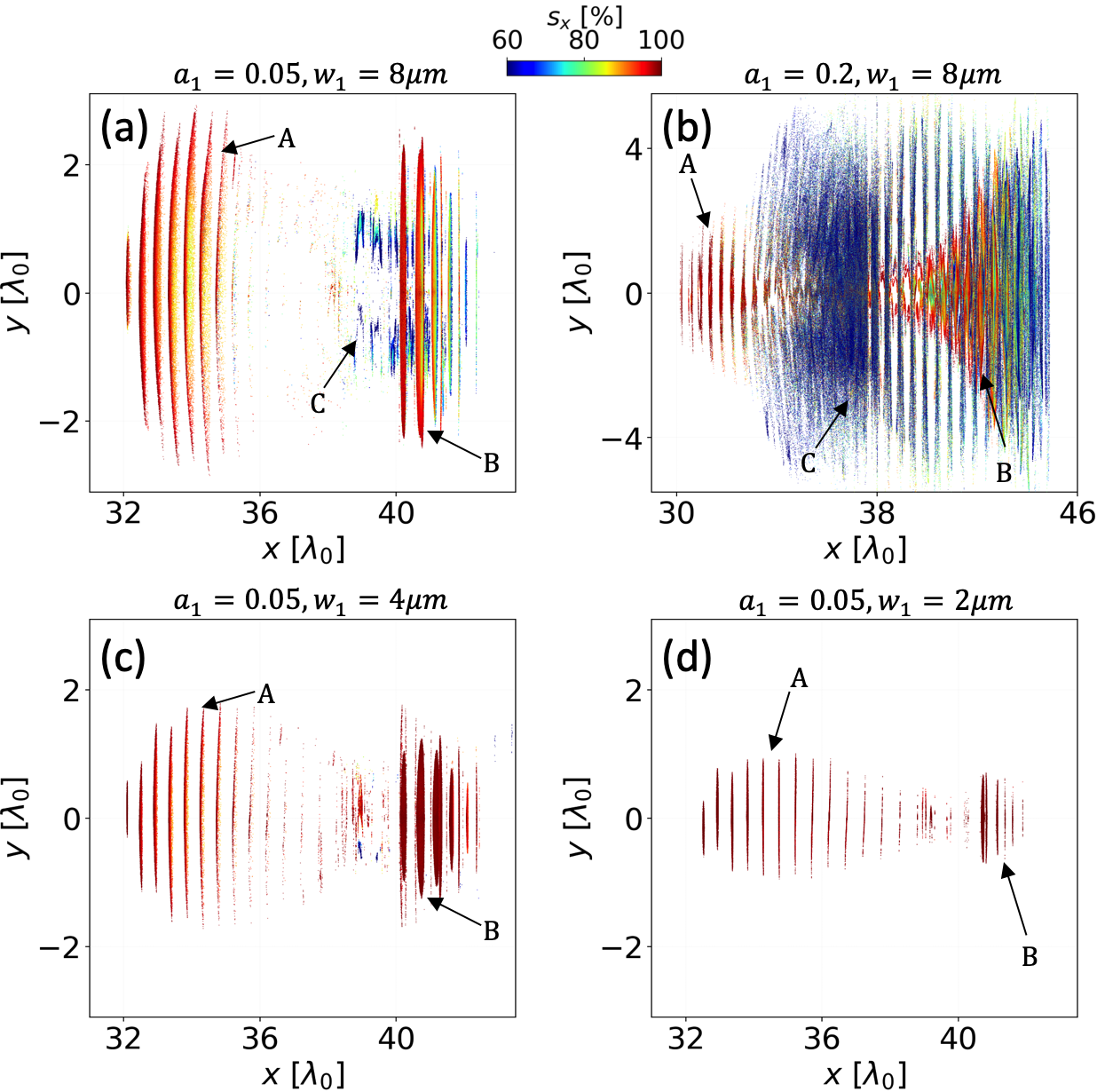}
\caption{FBPIC simulation results showing the initial distribution in the $(x,y)$ space of injected electrons for the same driving laser and plasma parameters as in Figs.~\ref{fig:fbpic3d}-\ref{fig:s_track} but for different colliding pulse parameters. (a) $a_1 = 0.05$ and $w_1 = \SI{8}{\micro\metre}$. (b) $a_1 = 0.2$ and $w_1 = \SI{8}{\micro\metre}$. (c) $a_1 = 0.05$ and $w_1 = w_0=\SI{4}{\micro\metre}$. (d) $a_1 = 0.05$ and $w_1 = \SI{2}{\micro\metre}$. The rainbow color denotes the electron longitudinal spin polarization $s_x$ at $t = 500\,T_0$.} \label{fig:w1}
\end{figure}

To test the validity of the scaling predicted by the injection criterion in Eq.~\eqref{eq:injection_criterion} in a more realistic three-dimensional scenario, we performed parametric scans with the spectral quasi-3D PIC code FBPIC~\cite{lehe2016spectral} with $N_m=2$ azimuthal modes, where $N_m>1$ accounts for departures from cylindrical symmetry in the fields. The simulation domain is a cylinder with a size of $x \times r = \SI{50}{\micro\metre} \times \SI{25}{\micro\metre}$, with cell size $\Delta x = 1/80\SI{}{\micro\metre}$ and $\Delta r = 1/40\SI{}{\micro\metre}$. Similarly to the 2D EPOCH simulations presented in Section~\ref{section II}, the two colliding laser pulses are linearly polarized along the $y$ axis with $\lambda_0 = \SI{0.8}{\micro\metre}$ wavelength, and have a Gaussian profile both in space and in time with $\tau_0 = \SI{25}{fs}$ FWHM of the intensity duration and $w_0=\SI{8}{\micro\metre}$ waist radius. The driving laser pulse normalized amplitude is $1 \leqslant a_0 \leqslant 2.5$ while the colliding laser pulse normalized amplitude is $0.01\leqslant a_1 \leqslant 0.2$. As in Section~\ref{section II} the plasma is underdense with a linearly growing density profile followed by a plateau with density $n_{e,0} = 10^{18}\,\mathrm{cm}^{-3}$ for $x > x_1$ [see Fig~\ref{fig:schematic}(a)]. The number of macro-particles per cell per species along the cylindrical coordinate axes $z$, $r$, and $\theta$ is $p_{n,z} = 4$, $p_{n, r} = 4$, and $p_{n,\theta} = 4$, respectively. 
\bl{Note that the spectral cylindrical representation implemented in FBPIC prevents numerical Cherenkov radiation (NCR) therefore permitting both fast and accurate simulations. To mitigate NCR, a finite difference cartesian 3D code would require a very small spatial and temporal time step, greatly increasing the computational cost. Simulations with $N_m=3$ did not show significant differences with respect to $N_m=2$, thus indicating that possible effects beyond cylindrical symmetry are properly accounted for. For simulations with laser pulses carrying a high orbital angular momentum, not considered in the present work, the inclusion of several azimuthal modes would be required.} \bl{In post-processing, the electron beam charge is self-consistently calculated from the weight of each macro electron. As the FBPIC simulation domain has a cylindrical geometry, the macroparticle weight $w$ is calculated as $w=n r d\theta dr dz$, where $n$ is the plasma number density, $r$ the radial coordinate, and $d\theta$, $dr$, and $dz$ are the grid steps along the azimuthal, radial, and longitudinal direction, respectively (see Ref.~\cite{FBPIC} for details on the FBPIC algorithm \& features).}

Figure~\ref{fig:injection} displays the model and FBPIC simulation results obtained with the above-mentioned parameters. In particular, Fig.~\ref{fig:injection}(a) and Fig.~\ref{fig:injection}(b) report the generated electron beam charge $\mathcal{Q}$ and average longitudinal spin polarization $\left\langle s_x \right\rangle$, respectively, while Fig.~\ref{fig:injection}(c) displays the average beam polarization as predicted by the model $\left\langle s_x \right\rangle = 1 - \kappa_s a_0 a_1$ (see Tab.~\ref{tab:sx_scaling}). The black dashed line in Fig.~\ref{fig:injection}(a)-(c) plots $a_1^*$, which is the injection threshold according to Eq.~\eqref{eq:injection_criterion}, where $\rho$ is used as a fitting parameter to the quasi-3D FBPIC simulations, which gives $\rho=0.55$. Remarkably, the scaling obtained from the model is in good agreement with the simulation results. Indeed, both the predicted scaling for electron injection as obtained in Eq.~\eqref{eq:injection_criterion}, which is set once $\rho$ is fixed, and the longitudinal beam polarization as estimated by assuming the simple scaling $\delta s_x \approx \kappa_s a_0^{m_0} a_1^{m_1}$ with the coefficients extracted from test particle simulations (see Tab.~\ref{tab:sx_scaling}), are in good agreement with simulations (see Fig.~\ref{fig:injection}).

Figure~\ref{fig:injection} shows that while highly-polarized beams are generated with laser field amplitudes chosen around the injection threshold $a_1 \approx a_1^*$, the electron beam charge $\mathcal{Q}$ is smaller for these lower field amplitudes. In fact, while the charge $\mathcal{Q}$ can be increased with higher $a_1$, this also results in a decrease of the average spin polarization $\left\langle s_x \right\rangle$ because of the stronger depolarization induced by the colliding laser pulses. By tuning the laser intensities $a_0$ and $a_1$, one can prioritize the electron charge $\mathcal{Q}$ or the spin polarization  of the beam. However, one of the key advantages of the CPI scheme is the availability of more degrees of freedom than LWFA with a single laser pulse. Such flexibility naturally lends itself to multiparameter space optimization, as shown in Ref.~\cite{bohlenInPrep}, where Bayesian optimization is employed to conceptually demonstrate high charge, highly polarized, and low emittance electron beam generation. In addition, the average charge delivered per unit time can be simply increased by employing relatively low power laser systems operating at high repetition rate.

Figure~\ref{fig:fbpic3d}(a)-(d) displays the electron density distribution, the transverse focusing force, the electron energy spectrum, and the average spin dependence on the energy of an electron beam obtained with $a_0 = 2$ and $a_1 = 0.05$. The corresponding laser intensity and power are $8.6 \times 10^{18}\text{ W/cm}^2$ and 8.7~TW, and $5.4 \times 10^{15}\text{ W/cm}^2$ and 5.4~GW for the driving and colliding pulse, respectively. Fig.~\ref{fig:fbpic3d}(a) shows that a bunch of electrons is injected at the rear of the first cavity. The transverse focusing force $-E_y+cB_z$ stabilizes electron acceleration in the cavity during the acceleration stage [Fig.~\ref{fig:fbpic3d}(b)]. The injected electron bunch has approximately 2.27~pC charge, 90\% average longitudinal spin polarization and about 33~MeV average energy after undergoing acceleration over $\approx \SI{450}{\micro\metre}$, with an accelerating gradient of approximately $90\,\mathrm{GeV m}^{-1}$. The electron energy spectrum $dN_e/d\varepsilon_e$ as a function of time clearly shows that the beam is highly monochromatic, and that the energy spread is preserved over the acceleration stage [see Fig.~\ref{fig:fbpic3d}(c)]. 
The average spin distribution $\left\langle s_x\right\rangle$ as a function of the electron energy $\varepsilon_e$ exhibits a correlation between the electron energy and their degree of polarization, with the higher energy electrons having predominantly a lower average spin [see Fig.~\ref{fig:fbpic3d}(d)]. 

Further insights can be gained by analyzing the temporal evolution of injected electrons [Fig.~\ref{fig:fbpic3d} panels~(e)] and by closely examining their distribution in the $(\xi, p_x)$ phase space [Fig.~\ref{fig:fbpic3d}(f)]. By comparing the electron distribution in $(\xi, p_x)$ at different times [see Fig.~\ref{fig:fbpic3d}(e) and the zoom of the distribution of injected electrons at $t=100\,T_0$ in Fig.~\ref{fig:fbpic3d}(f)], we find that the electrons with lower final spin polarization are injected with lower energy, initially, and with a longitudinal momentum near the injection threshold $p_{th}^-$ [see Fig.~\ref{fig:fbpic3d}(f)]. These electrons are later accelerated to higher energy inside the cavity as predicted by the Hamiltonian model [see the contours of the Hamiltonian in Figs.~\ref{fig:fbpic3d}(e)-(f)]. This explains the correlation between electron energy and average spin of electrons in the generated beam. The zoom of the electron $(\xi, p_x)$ distribution in Fig.~\ref{fig:fbpic3d}(f) suggests the existence of three distinct electron populations, which are labeled A, B, and C. This is corroborated by tracking the injected electrons back to their initial position and examining their evolution. Fig.~\ref{fig:fbpic3d}(g) plots the initial position of injected electrons in $(x, y)$, while Fig.~\ref{fig:fbpic3d}(h) displays the detailed evolution of the injected electrons in the longitudinal phase space $(x, p_x)$ for $t < 100 \, T_0$. Figures~\ref{fig:fbpic3d}(g)-(h) show that the electrons belonging to population~A interact with the driving laser pulse significantly before interacting with the colliding pulse, thus suggesting that the colliding pulse merely plays a perturbative role for population~A electrons. These electrons do not show a violent stochastic spin precession, and the spin polarization loss induced by the colliding pulses is insignificant.
Figure~\ref{fig:fbpic3d}(g) exhibits that the electrons belonging to populations~B and C first interact with the colliding laser pulse before experiencing the stronger fields of the driving pulse. Population~B and population~C electrons occupy nearly the same spatial region around $x\approx 40\,\lambda_0$, initially, and form transverse `stripes' that are longitudinally shifted by $\lambda_0/2$ [see Fig.~\ref{fig:fbpic3d}(g)]. Thus, population~B and population~C electrons have a phase difference $\phi_{1,B}-\phi_{1,C}=\pi$, which implies a different longitudinal momentum according to Eq.~\eqref{eq:px_py_colliding_l2}. As a result, while electrons belonging to population~B quickly slide away from the driving laser pulse and remain in the same region around $x\approx 40\,\lambda_0$, the electrons of population~C follow the driving pulse when they experience the combined fields of the colliding pulses, and drift up to $x\approx 45\,\lambda_0$ longitudinally [see the orange and green dots in Fig.~\ref{fig:fbpic3d}(h) corresponding to $t=70\,T_0$ and $t=90\,T_0$, respectively]. This relatively long interaction with the laser fields results in significant spin precession and depolarization. 

To confirm the above analysis, we have tracked the dynamics of two representative electrons extracted from each of the three populations~A, B, and C, with the two electrons having nearly the same initial conditions. Figure~\ref{fig:s_track}(a1) and Fig.\ref{fig:s_track}(a2) report the evolution of the three components of the momentum $p_{x}$, $p_{y}$, $p_{z}$ and of the spin $s_{x}$, $s_{y}$, $s_{z}$ of the representative electrons of population~A, respectively. Figure~\ref{fig:s_track}(a3) reports the evolution in $(ct-x, y)$ of the representative electrons of population~A. Figures~\ref{fig:s_track}(b1)-(b3) and \ref{fig:s_track}(c1)-(c3) report the same quantities for electrons extracted from population~B and C, respectively. Figure~\ref{fig:s_track} shows that the momentum and spin dynamics of electrons belonging to populations A and B is weakly dependent on initial conditions and, in particular, the electron spins nearly returns to their initial values after interaction [see Figs.~\ref{fig:s_track}(a2)-(b2)]. Notably, for populations A and B the electron transverse momentum $p_{y,z}$ closely follows the evolution predicted by Eq.~\eqref{eq:px_py_colliding_p} in the collision of two plane waves in vacuum, the anharmonic dynamics manifesting itself in the longitudinal momentum $p_x$, as expected from Eq.~\eqref{eq:px_py_colliding_l}. By contrast, electrons from population~C show signatures of premature and significant transverse momentum gain and a strong dependence on initial conditions. Unlike populations A and B, for population C the electron transverse momentum $p_{y,z}$ does not follow the evolution predicted by Eq.~\eqref{eq:px_py_colliding_p}, especially for $p_z$ [see Fig.~\ref{fig:s_track}(c1)]. This suggests that finite transverse size effects are important for this population. Note that the theoretical analysis of Section~\ref{section III} leverages test particle simulation results where the two colliding pulses are assumed to be plane waves in vacuum. Thus, deviations are expected if the waist radius of even one of the two laser pulses is not much larger than the laser wavelength.

A detailed study of how transverse effects and plasma self-generated fields influence the transverse momentum gain of the injected electrons is beyond the scope of the current work, and parameter optimization is carried out in a separate work~\cite{bohlenInPrep}. Nevertheless, it is worth noting that the contribution of electrons from population~C can be controlled by tuning the colliding laser pulse intensity and waist radius. Naturally, this affects both the total beam charge $\mathcal{Q}$ and its average spin polarization $\left\langle s_x \right\rangle$.
For instance, when the intensity of the colliding pulse is increased from $a_1 = 0.05$ [see Fig.~\ref{fig:w1}(a)] to $a_1 = 0.2$ [see Fig.~\ref{fig:w1}(b)] with all other parameters unchanged, the number of electrons from population~C greatly increases, and correspondingly the total charge of the electron beam rises from $\mathcal{Q} \approx 2.27\,$pC to $20.6\,$pC. Alternatively, the injection of electrons from population~C can be strongly suppressed by reducing the waist radius of the colliding laser pulse $w_1$. For instance, by reducing $w_1$ from $w_1 = \SI{8}{\micro\metre}$ to $w_1 = \SI{4}{\micro\metre}$ [see Fig.~\ref{fig:w1}(c)] and $w_1 = \SI{2}{\micro\metre}$ [see Fig.~\ref{fig:w1}(d)] and keeping all other parameters fixed, electrons from population~C are suppressed, while the average beam polarization is enhanced from $\left\langle s_x \right\rangle \approx 90.1\%$ to $\left\langle s_x \right\rangle \approx 95.3\%$ and $\left\langle s_x \right\rangle \approx 97.6\%$, respectively.

\section{\label{section V} Conclusion}

In conclusion, we have studied the dynamics of spin-polarized electron injection in the colliding-pulse scheme. By means of simple analytical modeling and multi-dimensional PIC simulations, we have shown that the electron injection process can be divided into a first stage of plasma electron collisionless heating and spin precession followed by a second stage of electron trapping and acceleration in the plasma wake. Leveraging test particle simulations and Hamiltonian analysis, we obtain a simple scaling for determining the electron injection threshold and the beam polarization as a function of the laser and plasma parameters. Further study is required to show the dependence on additional parameters, such as relative laser polarization. Model estimates are in good agreement with quasi-3D FBPIC simulations over a broad range of experimentally relevant laser parameters. While it was already shown that the colliding pulse injection scheme reliably provides electron beams with excellent quality~\cite{faureN06, kotaki2009electron, rechatin2009controlling, lundh2011few}, here we have shown that this scheme also enables control of the spin-polarization degree of the generated beam. Remarkably, the required relatively low laser power of this scheme, $a_0=2$ ($a_1=0.05$) and $w_0 = \SI{8}{\micro\metre}$ corresponding to 8.7~TW (5.4~GW), empower stable, reliable, and highly controllable operations even at high repetition rates, which is particularly relevant for applications such as precision measurements in fundamental physics~\cite{Tolhoek:1956:RevModPhys.28.277, QweakN2018}.

\begin{acknowledgments}
The original version of the PIC code EPOCH adapted here is funded by the UK EPSRC grants EP/G054950/1, EP/G056803/1, EP/G055165/1 and EP/ M022463/1. Z.~G. would like to thank Rong-Hao Hu for useful discussions. The authors gratefully acknowledge the Gauss Centre for Supercomputing e.V. (www.gauss-centre.eu) for providing computing time used for the FBPIC simulations through the John von Neumann Institute for Computing (NIC) on the GCS Supercomputer JUWELS at J\"ulich Supercomputing Centre (JSC).
\end{acknowledgments}

\section*{\NoCaseChange{Author Contributions}}
Z.G. carried out the simulations by using FBPIC with the spin dynamics model implemented by M.J.Q.. Z.G.performed the analysis with assistance from M.T.. The manuscript was written by Z.G. and M.T, with feedback from S.B. C.H.K. and K.P.. M.T. supervised the project. All authors discussed the results presented in the paper.

\appendix*

\section{Spin pusher}

The spin of an electron in an electric $\bd{E}$ and magnetic $\bd{B}$ field precesses according to the Thomas-Bargmann-Michel-Telegdi (TBMT) equation~\cite{thomas1927kinematics, bargmann1959precession}
\begin{equation} \label{eq:bmt}
\frac{\mathrm{d}\bd{s}}{\mathrm{d}t} = \bd{\Omega} \times \bd{s},
\end{equation}
where $\bd{\Omega} = \bd{\Omega_T} + \bd{\Omega_a}$ with
\begin{align}
\bd{\Omega_T} = & \frac{|e|}{m_e c}\left(\frac{\bd{B}}{\gamma} - \frac{\bd{\beta}}{1+\gamma}\times \bd{E} \right), \\
\bd{\Omega_a} = & \frac{a_e |e|}{m_e c}\left[\bd{B} - \frac{\gamma}{1+\gamma}\bd{\beta}(\bd{\beta}\cdot\bd{B}) - \bd{\beta}\times \bd{E} \right],
\end{align}
where $\gamma=\sqrt{1+\bd{p}^2/m_e^2 c^2}$ is the Lorentz factor of the electron, $\bd{\beta}=\bd{p}/\gamma m_e c$ is its normalized velocity, and $a_e \approx 1.16\times10^{-3}$ is the electron anomalous magnetic moment. Note the equations above are specific to electrons through their dependence on the anomalous magnetic moment $a_e$. The leap-frog equation obtained by discretizing Eq.~(\ref{eq:bmt}) and with the electromagnetic fields $\bd{E}^n$, $\bd{B}^n$ at step $n$ is
\begin{equation} \label{precession}
\frac{\bd{s}^{n+1/2} - \bd{s}^{n-1/2}}{\Delta t} =  \bd{\Omega}^n \times \bd{s}^n.
\end{equation}
Here we have used the following definition for the midpoint spin and momentum
\begin{align}
\bd{s}^n = & \frac{\left(\bd{s}^{n+1/2} + \bd{s}^{n-1/2}\right)}{2},\\
\bd{p}^n = & \frac{\left(\bd{p}^{n+1/2} + \bd{p}^{n-1/2}\right)}{2},\\
\gamma^n = & \sqrt{1+(\bd{p}^n)^2/m_e^2 c^2}.
\end{align}
By inserting these quantities into Eq.~(\ref{precession}) one immediately obtains $|\bd{s}^{n+1/2}| = |\bd{s}^{n-1/2}|$. Eq.~(\ref{precession}) can be rewritten as
\begin{equation} \label{preces2}
\bd{s}^{n+1/2} = \bd{s}' + (\bd{h} \times \bd{s}^{n+1/2}),
\end{equation}
where $\bd{h} = \bd{\Omega}^n \Delta t/2$ and $\bd{s}' = \bd{s}^{n-1/2} + \bd{h} \times \bd{s}^{n-1/2}$. Now, Eq.~(\ref{preces2}) is a linear system of equations in the unknown $\bd{s}^{n+1/2}$, whose solution is
\begin{equation} \label{eq:push_s}
\bd{s}^{n+1/2} = o \left[ \bd{s}' + (\bd{h} \cdot \bd{s}') \bd{h} + \bd{h} \times \bd{s}' \right],
\end{equation}
where $o = 1/ (1 + \bd{h}^2)$. The same approach discussed above can be employed for advancing the momentum \cite{tamburini11}.

Following the work in \bl{Refs.~\onlinecite{wen2019polarized, H2019Polarized}}, several schemes leveraging pre-polarized plasma generation via laser-induced molecular photodissociation~\cite{rakitzis2003spin_science, sofikitis2017spin_PRL, sofikitis2018spin_PRL} have been proposed~\cite{wu2019polarized_NJP, wu2019polarized_PRE, Buscher_2019, jin2020spin, gong2020energetic, thomas2020scaling, li2021polarized, reichwein2021robustness, reichwein2022acceleration, reichwein2022spin, reichwein2022particle, Fan_2022_njp, yan2022generation}. Although not used in the present article, we implemented in FBPIC not only the electron spin degrees of freedom and their evolution as detailed above, but also the capability to \bl{approximately} model the initial spin state of electrons ionized from the photodissociation products of hydrogen halide molecules such as HCl. In fact, as with the momentum and position, an initial value for the spin must be provided for all species of particles.

For molecules such as HCl, which are not considered in this paper, the unpaired outer-shell electron of H and Cl has an initial spin along the propagation axis of the dissociation laser after ionization, whereas the many spin-paired inner-shell electrons must be treated differently. In practice, when an inner shell electron of Cl is ionized, its initial spin is randomly oriented in space, to account of the fact that no orientation is present for these electrons. Ideally, a more sophisticated model would include quantum mechanical effects when determining the initial spin of each successive electron emitted by ionization, but our simpler approach suffices for capturing the essential dynamics of polarized electrons obtained with the technique of laser-induced molecular photodissociation.

%

\end{document}